\documentclass[epj]{svjour}
\renewcommand{\bar}[1]{\overline{#1}}
\renewcommand{\d}{{\mathrm d}}

\usepackage{graphics}
\begin{document}
\title{Fragmentation functions of mesons in the Field-Feynman
model}
\author{Jing Hua\inst{1} \and
Bo-Qiang Ma\inst{2}$^{,}$\thanks{Email address: mabq@phy.pku.edu.cn; corresponding author. }
}                     
%
%
\institute{Department of Physics, Peking University, Beijing
100871, China \and CCAST (World Laboratory), P.O.~Box 8730,
Beijing 100080, China\\ Department of Physics, Peking University,
Beijing 100871, China}
%
\date{Received: date / Revised version: date}
%
\abstract{ The fragmentation functions of the pion with
distinction between $D_{u}^{\pi^{+}}$, $D_{d}^{\pi^{+}}$, and
$D_{s}^{\pi^{+}}$ are studied in the Field-Feynman recursive
model, by taking into account the flavor structure in the
excitation of quark-antiquark pairs by the initial quarks. The
obtained analytical results are compatible with available
empirical results. The framework is also extended to predict the
fragmentation functions of the kaon with distinction between
$D_{\bar{s}}^{K^{+}}(z)$, $D_{u}^{K^{+}}(z)$, $D_{s}^{K^{+}}(z)$,
and $D_{d}^{K^{+}}(z)$. This work gives a significant modification
of the original model, and the predictions can be tested by future
experiments on the fragmentation functions of the kaon.
\PACS{
      {13.87.Fh}{}   \and
      {12.38.Lg}{}   \and
      {13.60.Le}{}   \and
      {14.40.Aq}{}
     } 
} 
\maketitle

\section{Introduction}

The productions of mesons are under investigations in various
processes nowadays, therefore the fragmentation functions of
mesons are among the useful basic quantities in high energy
physics studies. A complete knowledge of the quark to meson
fragmentation functions with detailed flavor structure is needed
and will be more useful. In previous studies, the fragmentation
functions of mesons are usually obtained from experimental data on
$e^+e^-$ annihilation process \cite{PDG}. However, the situation
has now been changed because the HERMES collaboration has
published results of charge separated data for $\pi^{\pm}$
production on the proton target \cite{HERMES}. By combining the
HERMES data on semi-inclusive DIS (SIDIS) $\pi^{\pm}$-production
with the singlet fragmentation function $D_{\Sigma}^{\pi^{+}}$,
which is well determined from $e^{+}e^{-}$ data, Kretzer, Leader,
and Christova \cite{KLC} obtained a complete set of fragmentation
functions of the pion: $D_{u}^{\pi^{+}}$, $D_{d}^{\pi^{+}}$, and
$D_{s}^{\pi^{+}}$, without specific assumption about relation
between favored and unfavored transitions, such as
$D_{d}^{\pi^{+}}(z)=(1-z)D_{u}^{\pi^{+}}(z)$ in \cite{K00} and
$D_{d}^{\pi^{+}}(z)=(1-z)^{n}D_{u}^{\pi^{+}}(z)$ (with n=2,3,4) in
\cite{MSY}. The Kretzer-Leader-Christova (KLC) parametrization
results of the experimental data are:
\begin{equation}
D_{u}^{\pi^{+}}(z)=0.689z^{-1.039}(1-z)^{1.241}, \label{KLC1}
\end{equation}
\begin{equation}
D_{d}^{\pi^{+}}(z)=0.217z^{-1.805}(1-z)^{2.037}, \label{KLC2}
\end{equation}
\begin{equation}
D_{s}^{\pi^{+}}(z)=0.164z^{-1.927}(1-z)^{2.886}. \label{KLC3}
\end{equation}

Although the evolution behaviors can be calculated perturbatively,
the actual form of the fragmentation functions is
non-perturbative. Therefore the fragmentation functions are
helpful to elucidate the fundamental features of hadronization.
The pioneering analysis by Field and Feynman \cite{FF,Field}
provides a simple and lucid picture to understand the
fragmentation for a quark into mesons based on a recursive
principle, and has stimulated more sophisticated models like the
string fragmentation model, implemented in Monte Carlo generation
programs \cite{PDG}, as in the Lund model \cite{Lund}. The
Field-Feynman model is successful to parameterize the
fragmentation functions of the pion and kaon with only two
parameters. However, no distinction between $D_{d}^{\pi^{+}}(z)$
and $D_{s}^{\pi^{+}}(z)$ was made in the original model \cite{FF},
for the sake to reduce independent parameters. With the recent
progress of clear distinction between $D_{d}^{\pi^{+}}(z)$ and
$D_{s}^{\pi^{+}}(z)$, it becomes mature and necessary to consider
the flavor structure of the unfavored fragmentation functions of
mesons in the Field-Feynman model. The purpose of this paper is,
for the first time, to make a distinction between
$D_{d}^{\pi^{+}}(z)$ and $D_{s}^{\pi^{+}}(z)$ in the Field-Feynman
model. It will be shown that this can be realized by adopting the
flavor dependent parameters for the excitation of sea
quark-antiquark pairs by the initial quarks. We will find that the
obtained analytical results for $D_{u}^{\pi^{+}}$,
$D_{d}^{\pi^{+}}$, and $D_{s}^{\pi^{+}}$ with only three
parameters, are compatible with the recent KLC parametrization
Eqs.~(\ref{KLC1}-\ref{KLC3}) based on experimental results. We
will also extend the same framework to the fragmentation functions
of the kaon, and show that we can distinguish between
$D_{\bar{s}}^{K^{+}}(z)$, $D_{u}^{K^{+}}(z)$, $D_{s}^{K^{+}}(z)$,
and $D_{d}^{K^{+}}(z)$ without introducing any additional
parameters. Thus the predictions can be tested by future
experiments on the fragmentation functions of the kaon.

The paper is organized as follows. In Sec.~2, we present a brief
review of the Field-Feynman recursive model for the quark to meson
fragmentation functions, and make a distinction between the
parameters that can introduce the difference between
$D_{d}^{\pi^{+}}(z)$ and $D_{s}^{\pi^{+}}(z)$. In Sec.~3, we
adjust the parameters based on the KLC parametrization and the
experimental results. We present the analytical results of
$D_{u}^{\pi^{+}}$, $D_{d}^{\pi^{+}}$, and $D_{s}^{\pi^{+}}$, and
compare them with the experimental results. In Sec.~4, we extend
the framework to the fragmentation functions of the kaon, and make
predictions of the four different fragmentation functions
$D_{\bar{s}}^{K^{+}}(z)$, $D_{u}^{K^{+}}(z)$, $D_{s}^{K^{+}}(z)$,
and $D_{d}^{K^{+}}(z)$. Finally, we present a summary in Sec.~5.

\section{A brief review of the recursive model}

\begin{figure}[htb]
\centering
\scalebox{0.8}[0.8]{\includegraphics*[148pt,582pt][403pt,772pt]{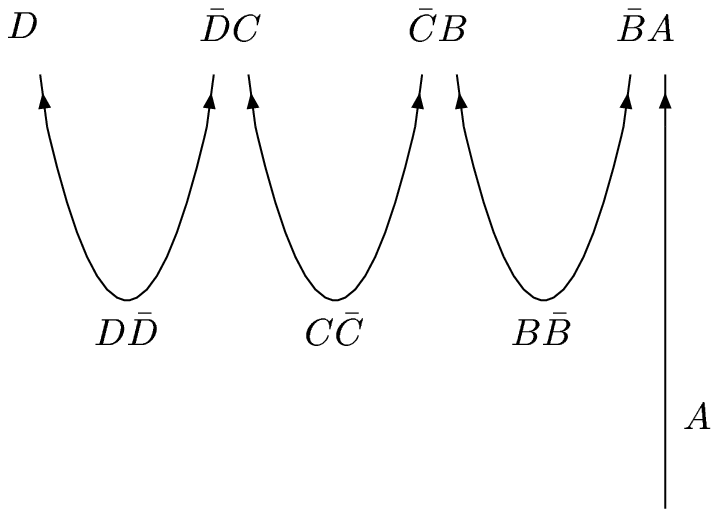}}
\caption[*]{\small Illustration of the Field-Feynman recursive
model for the fragmentation of mesons. The initial quark $A$
combines with $\bar{B}$ to form the rank 1 ``primary" meson with
configuration $\bar{B}A$, and the remained quark $B$ goes on the
same way as the former quark $A$. }\label{f1}
\end{figure}

The Field-Feynman model \cite{FF} of the meson fragmentation is on
the basis of a recursive principle, as can be illustrated in
Fig.~\ref{f1}. The model ansatz is based on the idea that an
incoming quark ``$A$" combines with an antiquark ``$\bar{B}$" from
a quark-antiquark pair ``$B\bar{B}$" which is produced from the
color field excited by the incoming quark ``$A$". Then the meson
``$A\bar{B}$" is constructed, and is ranked as 1st primary meson.
The remaining quark ``$B$" goes on the same way as the quark
``$A$", and combines with ``$\bar{C}$" from a quark-antiquark pair
``$C\bar{C}$" which is excited by the quark ``$B$".  It constructs
the primary meson ``$B\bar{C}$", ranked as 2nd primary meson. Then
the remaining quark ``$C$"  goes on the same way in the remaining
cascade.

With $f(\eta)\d \eta$ denoting the probability that the first rank
1 meson leaves a fractional momentum $\eta$ to the remaining
cascade, the function $f(\eta)$ is normalized by
\begin{equation}
\int_{0}^{1}f(z)\d z =1 .
\end{equation}
Then let $F(z)\d z$ be the probability of finding a meson
(independent of rank) with fractional momentum $z$ within $\d z$
in a quark jet. So $F(z)$ can be written by the recursive
principle
\begin{equation}
F(z)=f(1-z)+\int_{0}^{1}\frac{\d
\eta}{\eta}f(\eta)F(\frac{z}{\eta}). \label{IE}
\end{equation}
The first term comes from the ranked 1 primary meson. The second
term calculates the probability of the production of a higher rank
meson recursively. Field and Feynman \cite{FF} found a solution
for the above integral equation with
\begin{equation}
zF(z)=f(1-z),
\end{equation}
by choosing
\begin{equation}
f(z)\equiv(d+1)z^{d}.
\end{equation}

With $\beta_{ij}$ being the probability of finding the
$q_{j}\bar{q}_{j}$ pair in the quark sea excited by the initial
quark $i$, the normalization condition imposes that
\begin{equation}
\sum_{j=1}^{n_{f}}\beta_{ij}=1.
\end{equation}
Here, we add an $i$ index which does not appear in the original
model, and we will find that it is useful. The modification of
this model is based on the difference between $\beta_{uu}$ and
$\beta_{su}$, which were assumed to be the same in the original
Field-Feynman model for the sake to reduce independent parameters.
The original model predicts no distinction between
$D_{d}^{\pi^{+}}$ and $D_{s}^{\pi^{+}}$, and the results in
\cite{KLC} show that this is not the realistic situation. Adopting
SU(2) symmetry between $u$ and $d$ quarks
\begin{equation}
\beta_{iu}=\beta_{id}\equiv \beta_{i},
\end{equation}
we get
\begin{equation}
\beta_{is}=1-2\beta_{i}.
\end{equation}
For an initial quark of flavor $q$, the mean number of fragmented
mesons with configuration $a\bar{b}$ at $z$ is given, in analogy
of Eq.~(\ref{IE}), by
\begin{equation}
P_{q}^{a\bar{b}}(z)=\delta_{aq}\beta_{qb}f(1-z)+\int_{z}^{1}\frac{\d
\eta}{\eta}f(\eta)\beta_{qc}P_{c}^{a\bar{b}}(z/\eta). \label{PIE1}
\end{equation}
The mean number of meson states averaged over all quarks is
\begin{equation}
P_{<q>}^{a\bar{b}}(z)=\sum_{c}\beta_{qc}P_{qc}^{a\bar{b}}(z),
\end{equation}
so that
\begin{equation}
P_{<q>}^{a\bar{b}}(z)=\beta_{qa}\beta_{qb}f(1-z)+\int_{z}^{1}\frac{\d
\eta}{\eta}f(\eta)\beta_{qc}P_{<c>}^{a\bar{b}}(z/\eta).
\label{PIE2}
\end{equation}
Comparing with Eq.~(\ref{IE}), one yields
\begin{equation}
P_{<q>}^{a\bar{b}}(z)=\beta_{qa}\beta_{qb}F(z).
\end{equation}
Substituting it back into Eq.~(\ref{PIE2}), one yields
\begin{equation}
P_{q}^{a\bar{b}}(z)=\delta_{aq}\beta_{qb}f(1-z)+\beta_{qa}\beta_{qb}\overline{F}(z),
\end{equation}
where
\begin{equation}
\begin{array}{ll}
\overline{F}(z) &\equiv
F(z)-f(1-z)=(\frac{1}{z}-1)f(1-z)\\&=(d+1)z^{-1}(1-z)^{d+1}.
\end{array}
\end{equation}
The fragmentation function is
\begin{equation}
D_{q}^{h}(z)=\sum_{ab}\Gamma_{a\bar{b}}^{h}P_{q}^{a\bar{b}}(z),
\end{equation}
where $\Gamma_{a\bar{b}}^{h}$ is the probability of a meson $h$
with configuration $a\bar{b}$. For example,
$\Gamma_{u\bar{d}}^{\pi^{+}}=1$ and
$\Gamma_{u\bar{u}}^{\pi^{0}}=\Gamma_{d\bar{d}}^{\pi^{0}}=1/2$.
Combining Eqs.~(\ref{PIE1}) and (\ref{PIE2}), we obtain
\begin{equation}
D_{q}^{h}(z)=A_{q}^{h}f(1-z)+B_{q}^{h}\overline{F}(z),
\end{equation}
where
\begin{equation}
A_{q}^{h}=\sum_{b}\Gamma_{q\bar{b}}^{h}\beta_{qb},
\end{equation}
\begin{equation}
B_{q}^{h}=\sum_{a,b}\beta_{qa}\Gamma_{a\bar{b}}^{h}\beta_{qb}.
\end{equation}
Then the fragmentation functions of $D_{u}^{\pi^{+}}$,
$D_{d}^{\pi^{+}}$, and $D_{s}^{\pi^{+}}$ can be written as
\begin{equation}
D_{u}^{\pi^{+}}(z)=\beta_{u}f(1-z)+(\beta_{u})^{2}\overline{F}(z),
\label{HMffp1}
\end{equation}
\begin{equation}
D_{d}^{\pi^{+}}(z)=(\beta_{u})^{2}\overline{F}(z), \label{HMffp2}
\end{equation}
\begin{equation}
D_{s}^{\pi^{+}}(z)=(\beta_{su})^{2}\overline{F}(z)=(\beta_{s})^{2}\overline{F}(z),
\label{HMffp3}
\end{equation}
here we adopt the notation $\beta_s=\beta_{su}$.

In the original Field-Feynman model, the parameters $\beta_{u}$
and $\beta_{s}$ are identical: $\beta_{u}=\beta_{s}$. This means
that the effect of producing quark-antiquark pairs excited by the
incident $s$ or $u$ quark is assumed to be the same. This leads to
the result of $D_{d}^{\pi^{+}}(z)=D_{s}^{\pi^{+}}(z)$. Indeed, the
effect due to the quark mass in the hadronization process strictly
speaking cannot be ignored. In other words, the excitation of a
quark-antiquark pair by incident quarks with different masses will
be different. This can be taken into account by introducing
$\beta_u \neq \beta_s$, and thus obtain $D^{\pi^+}_u(z)\neq
D^{\pi^+}_u(z)$. That is, we take into account the difference in
the excitation of quark-antiquark pairs by the initial
light-flavor ($u$ or $d$) and strange ($s$) quarks. Here, $SU(2)$
symmetry is still assumed, $\beta_{qu}= \beta_{qd}=\beta_{q}$.

\section{The fragmentation functions of the pion}

From (\ref{HMffp1}) and (\ref{HMffp2}), two relations can be
obtained
\begin{equation}
D_{d}^{\pi^{+}}(z)/D_{u}^{\pi^{+}}(z)=\frac{\beta_{u}(1/z-1)}{(1-\beta_{u}+\beta_{u}/z)},
\end{equation}

\begin{equation}
D_{u}^{\pi^{+}}(z)-D_{d}^{\pi^{+}}(z)=\beta_{u}f(1-z).
\label{Du-Dd}
\end{equation}

\begin{figure}
\resizebox{0.5\textwidth}{!}{%
  \includegraphics{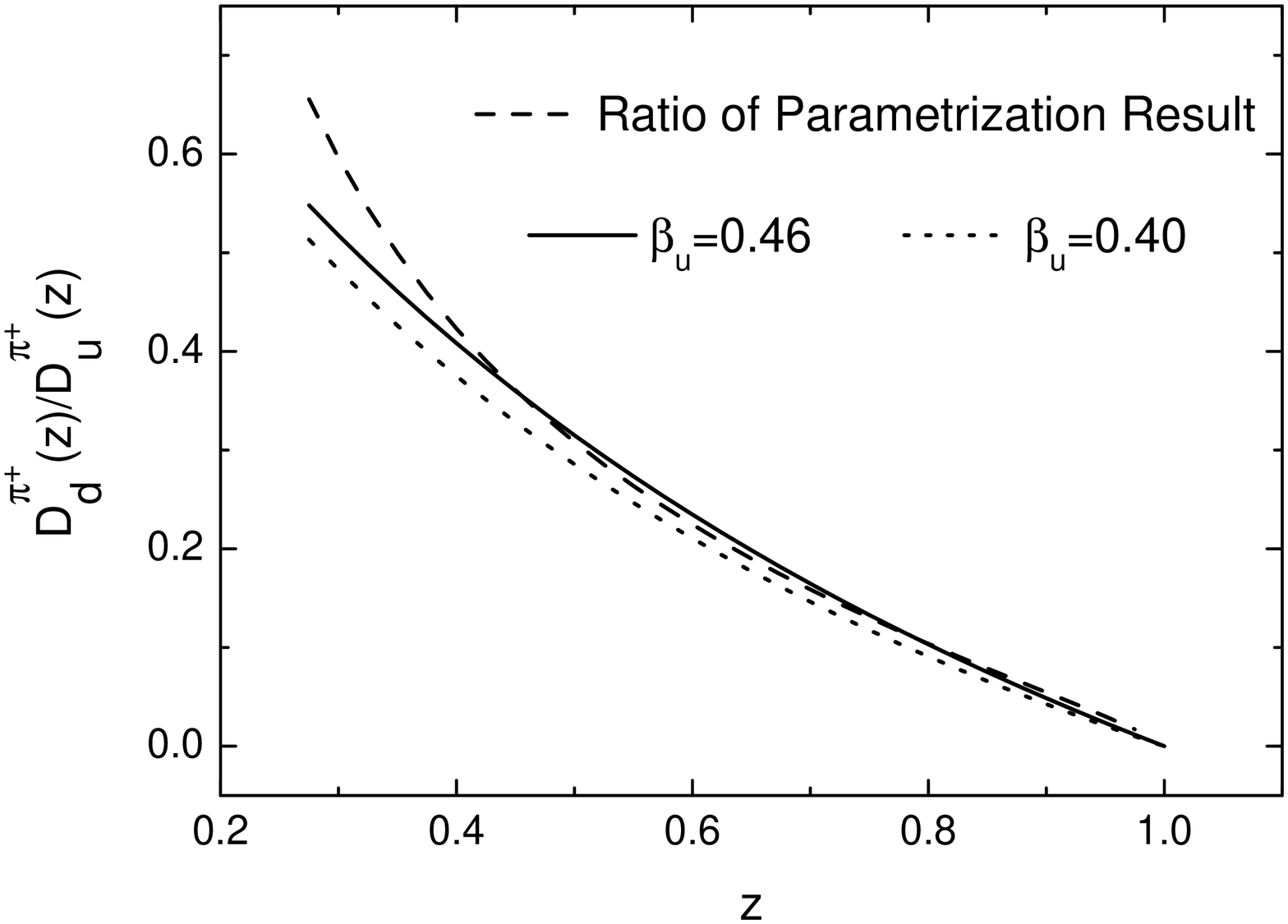}}
\vspace{0.1cm}
 \caption{\small The dash curve is the ratio of
$D_{d}^{\pi^{+}}/D_{u}^{\pi^{+}}$ from the KLC parametrization
\cite{KLC}, and the solid and dot curves are the analytical
results of this work with two different values of $\beta_{u}$.
}\label{f2}
\end{figure}

\begin{figure}
\resizebox{0.5\textwidth}{!}{%
  \includegraphics{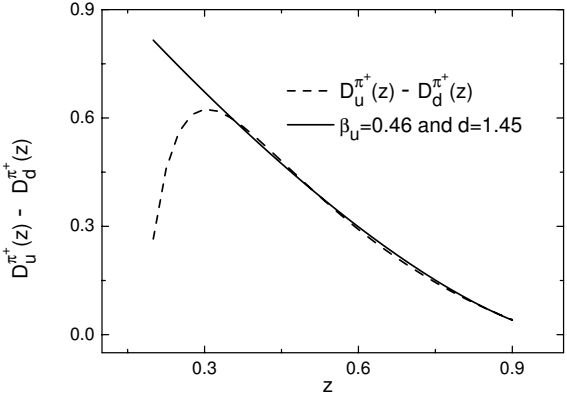}}
\caption{\small The dash curve is
$D_{u}^{\pi^{+}}-D_{d}^{\pi^{+}}$ from the KLC parametrization
\cite{KLC}, and the solid curve is $\beta_{u}(d+1)(1-z)^{d}$
(Eq.~(\ref{Du-Dd})) with $\beta_{u}=0.46$ and $d=1.45$.
}\label{f3}
\end{figure}

\begin{figure}
\resizebox{0.5\textwidth}{!}{%
  \includegraphics{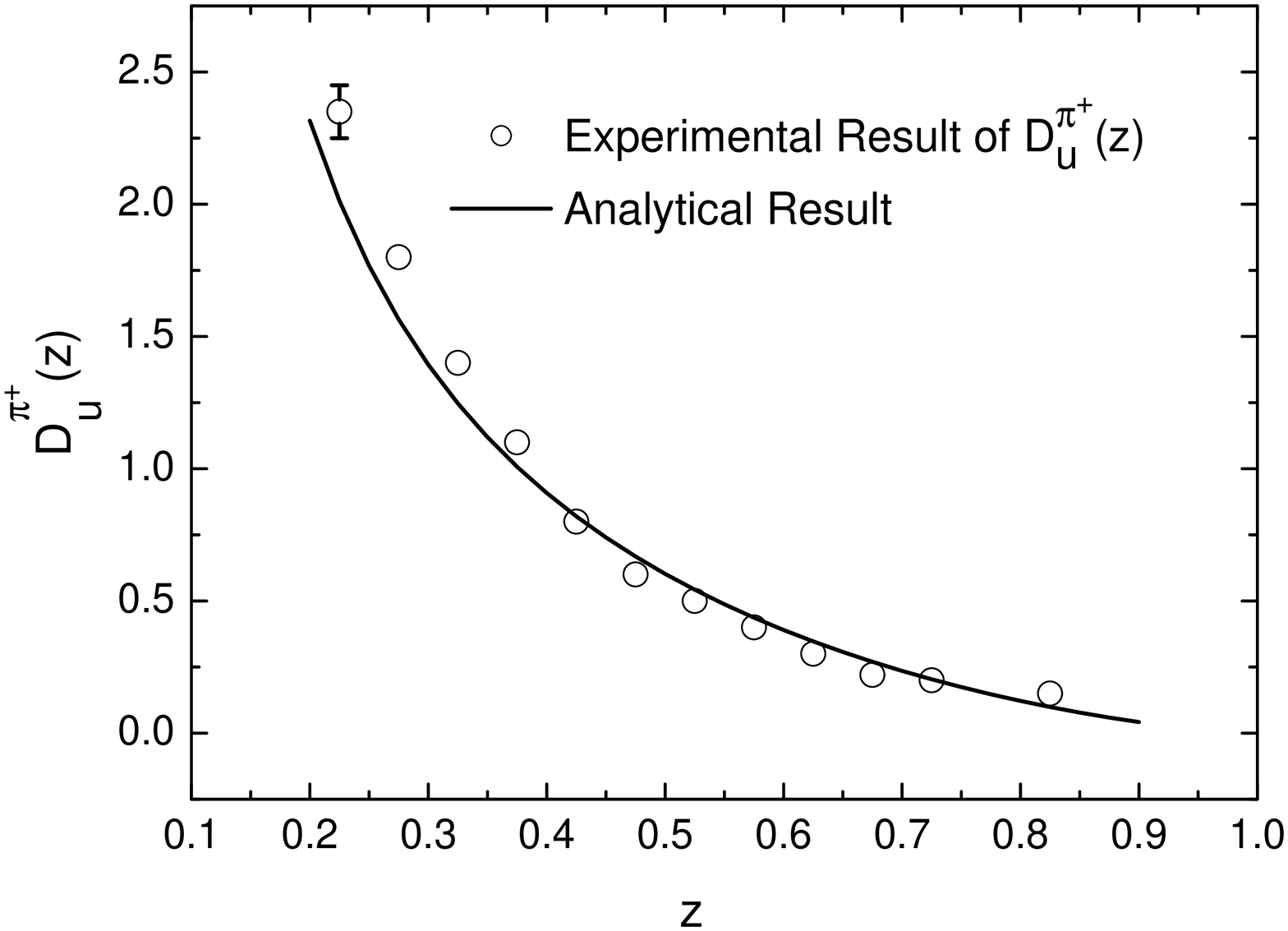}}
\caption{\small The circles are the experimental results from
\cite{KLC}, and the solid curve is the analytical result of this
work. }\label{f4}
\end{figure}

\begin{figure}
\resizebox{0.5\textwidth}{!}{%
  \includegraphics{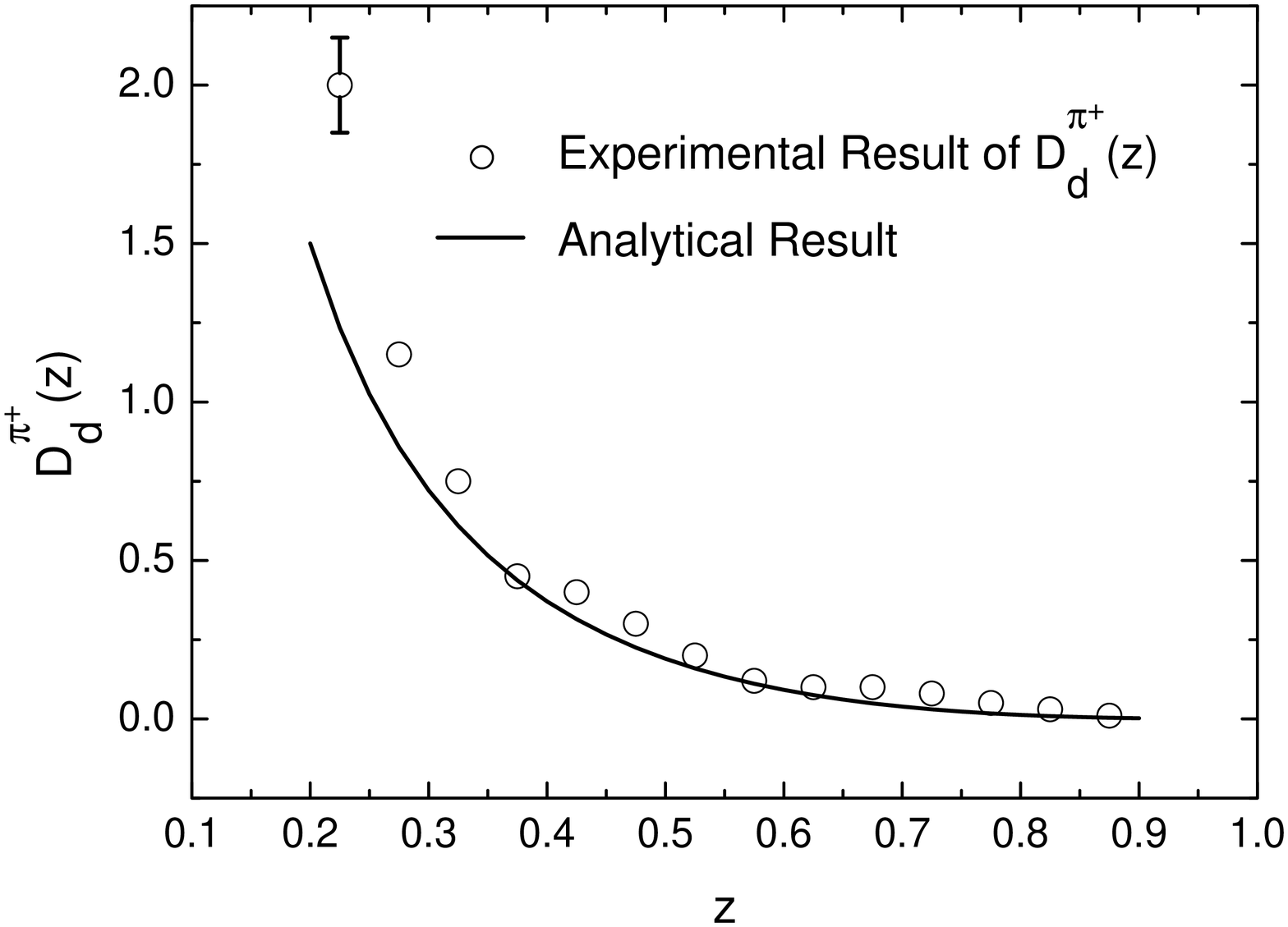}}
\caption{\small The circles are the experimental results from
\cite{KLC}, and the solid curve is the analytical result of this
work. }\label{f5}
\end{figure}

The first relation is independent of $f(z)$, so that it is useful
to set the parameter $\beta_{u}$. The parametrization results
\cite{KLC} of $D_{u}^{\pi^{+}}(z)$ and $D_{d}^{\pi^{+}}(z)$ are
used to adjust the parameters. We try to set the value of
$\beta_{u}$ by plotting $D_{d}^{\pi^{+}}(z)/D_{u}^{\pi^{+}}(z)$
for the analytical result in this work and the parametrization
result by KLC, as shown in Fig.~\ref{f2}. We find that
$\beta_{u}=0.46$ in the domain of $z \in (0.3,1)$. We also adopt
the option of $\beta_{u}=0.4$, which was used in the original
model \cite{FF}, as a comparison.  Then plotting the function
$D_{u}^{\pi^{+}}(z)-D_{d}^{\pi^{+}}(z)$ in this work and in the
KLC parametrization, we can get $d=1.45$ with $\beta_{u}=0.46$, as
shown in Fig.~\ref{f3}. It is surprising that we can get
compatible curves by only adjusting the parameters $\beta_{u}$ and
$d$ at $z>0.3$, as $f(z)=(d+1)z^{d}$ is only an assumption in the
original model. There is no general reason for this choice except
for the feasibility of solving the integral equation (\ref{IE}) in
an exact way. And now, at some degree, the KLC parametrization
results extracted from experimental results without any assumption
about favored and unfavored transitions, support the Field-Feyman
assumption of $f(z)$ as quite right in the domain of $z\in
(0.3,1)$. Substituting $\beta_{u}=0.46$ and $d=1.45$ into
(\ref{HMffp1}) and (\ref{HMffp2}), we get the analytical results
of $D_{u}^{\pi^{+}}(z)$ and $D_{d}^{\pi^{+}}(z)$,
\begin{equation}
D_{u}^{\pi^{+}}(z)=1.127(0.54 +0.46/z)(1-z)^{1.45},
\label{HMffp1n}
\end{equation}
\begin{equation}
D_{d}^{\pi^{+}}(z)=0.51842z^{-1}(1-z)^{2.45}. \label{HMffp2n}
\end{equation}
We compare the analytical results of (\ref{HMffp1n}) and
(\ref{HMffp2n}) with the experimental results in Fig.~\ref{f4} and
Fig.~\ref{f5}. It is obvious that the analytical results of
(\ref{HMffp1n}) and (\ref{HMffp2n}) can well explain the
experimental results, especially in the domain of $z\in (0.3,1)$.

\begin{figure}
\resizebox{0.5\textwidth}{!}{%
  \includegraphics{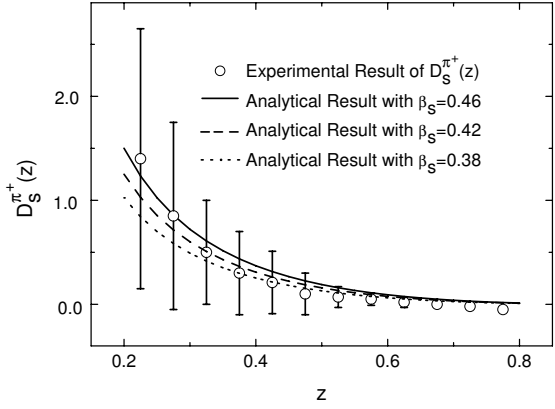}}
\vspace{0.08cm}
 \caption{\small The circles are the experimental
results from \cite{KLC}, and the three curves show three
analytical results of this work with different $\beta_{s}$. It
goes back to the original Field-Feynman model when
$\beta_{s}=\beta_{u}=0.46$. }\label{f6}
\end{figure}

We now look at the unfavored fragmentation function
$D_{s}^{\pi^{+}}(z)$, which was predicted to be the same as the
unfavored fragmentation function $D_{d}^{\pi^{+}}(z)$ in the
original model. We present the experimental results of
$D_{s}^{\pi^{+}}(z)$ and the analytic result (\ref{HMffp3}) of
this work in Fig.~\ref{f6}, with the value of $\beta_{s}$ adjusted
to be $0.38$. That means that $38\%$ quark-antiquark pairs in the
color field excited by an $s$ quark are the light-flavor
$u\bar{u}$ or $d\bar{d}$ pairs. Whereas there are $46\%$
$u\bar{u}$ or $d\bar{d}$ quark-antiquark pairs excited by a
light-flavor $u$ or $d$ quark. We should notice that
$\beta_{ss}\neq \beta_{us}$, as $\beta_{ss}=1-2
\beta_{su}=1-2\beta_s=0.24$ is the fraction of $s\bar{s}$ pairs
excited by an $s$ quark, whereas $\beta_{us}=1-2 \beta_{uu}=1-2
\beta_u=0.08$ is the fraction of $s\bar{s}$ pairs excited by a $u$
or $d$ quark. This is quite reasonable, considering that there
should be more fraction of $s\bar{s}$ pairs excited by an $s$
quark than that by a light-flavor ($u$ or $d$) quark. Therefore we
have
\begin{equation}
D_{s}^{\pi^{+}}(z)=0.35378z^{-1}(1-z)^{2.45}.
\end{equation}
If we choose $\beta_{s}=0.46$, $D_{s}^{\pi^{+}}(z)$ is identical
to $D_{d}^{\pi^{+}}(z)$, and we return back to the original model.

\begin{figure}
\resizebox{0.5\textwidth}{!}{%
  \includegraphics{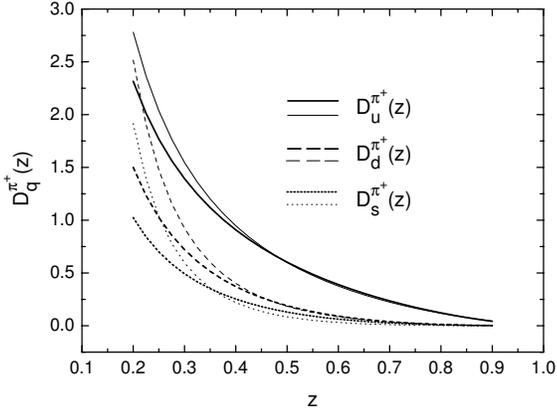}}
\caption{\small The fragmentation functions of the pion with
distinction between $D_{u}^{\pi^{+}}$, $D_{d}^{\pi^{+}}$, and
$D_{s}^{\pi^{+}}$, The thick curves are the analytical results of
this work, and the thin curves are the KLC parametrization results
\cite{KLC}. }\label{f7}
\end{figure}

We present in Fig.~\ref{f7} our analytical results for a complete
set of fragmentation functions of the pion with distinction
between $D_{u}^{\pi^{+}}$, $D_{d}^{\pi^{+}}$, and
$D_{s}^{\pi^{+}}$, and compare them with the KLC parametrization
results. We find that the two sets of fragmentation functions are
compatible with each other, especially at $z>0.3$. Considering
that a large uncertainty may exist in the parametrization of the
unfavored fragmentation function $D_{d}^{\pi^{+}}(z)$ at small
$z$, we may consider the analytical results in this work as an
alternative set of fragmentation functions for possible
applications, for example, to check the sensitivity on different
fragmentation functions in the predictions of various pion
fragmentation processes \cite{MSY}.

One notices from Figs.~\ref{f2} and \ref{f3} that the analytical
results differ by those obtained by KLC in the small $z$ region,
as also happens for the analytical results and the experimental
data in Figs.~\ref{f4} and \ref{f5}. This feature is easy to be
understood, as the sea content of the meson has not been taken
into account in the Field-Feynman model. The small $z$ behaviors
of the fragmentation functions are also influenced by the sea
quark-antiquark pairs inside the meson \cite{MSSY}. Thus it is
necessary to consider this aspect if one wishes to describe the
small $z$ behaviors of fragmentation functions with a more
reasonable picture.

\section{The fragmentation functions of the kaon}

We now apply the same framework to the fragmentation functions of
the kaon. Following the same procedures, we can get
\begin{equation}
D_{\bar{s}}^{K^{+}}(z)=\beta_{s}f(1-z)+\beta_{s}(1-2\beta_{s})\overline{F}(z),
\end{equation}
\begin{equation}
D_{u}^{K^{+}}(z)=(1-2\beta_{u})f(1-z)+\beta_{u}(1-2\beta_{u})\overline{F}(z),
\end{equation}
\begin{equation}
D_{s}^{K^{+}}(z)=\beta_{s}(1-2\beta_{s})\overline{F}(z),
\end{equation}
\begin{equation}
D_{d}^{K^{+}}(z)=D_{\bar{u}}^{K^{+}}(z)=D_{\bar{d}}^{K^{+}}(z)=\beta_{u}(1-2\beta_{u})\overline{F}(z),
\end{equation}
where $\beta_{s}=\beta_{su}$, $\beta_{u}=\beta_{uu}$, and
$\beta_{qs}=1-2\beta_{q}$.

\begin{figure}[htb]
\resizebox{0.5\textwidth}{!}{%
  \includegraphics{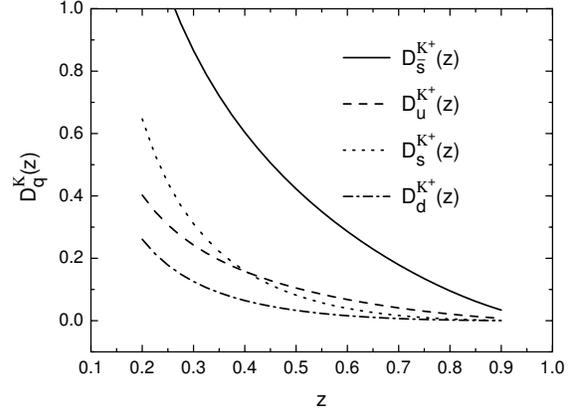}}
\caption{\small The fragmentation functions of the kaon in this
work with distinction between $D_{\bar{s}}^{K^{+}}(z)$,
$D_{u}^{K^{+}}(z)$, $D_{s}^{K^{+}}(z)$, and $D_{d}^{K^{+}}(z)$.
}\label{f8}
\end{figure}

With the parameters $\beta_{u}=0.46$, $\beta_{s}=0.38$, and
$d=1.45$ being the same as for the pion,  we get a complete set of
fragmentation functions of the kaon
\begin{equation}
D_{\bar{s}}^{K^{+}}(z)=0.931(0.76+0.24/z)(1-z)^{1.45},
\end{equation}
\begin{equation}
D_{u}^{K^{+}}(z)=0.196(0.54+0.46/z)(1-z)^{1.45},
\end{equation}
\begin{equation}
D_{s}^{K^{+}}(z)=0.22344z^{-1}(1-z)^{2.45},
\end{equation}
\begin{equation}
D_{d}^{K^{+}}(z)=0.09016z^{-1}(1-z)^{2.45}.
\end{equation}
These analytical results of the fragmentation functions of the
kaon are plotted in Fig.~\ref{f8} with four independent curves:
$D_{\bar{s}}^{K^{+}}(z)$, $D_{u}^{K^{+}}(z)$, $D_{s}^{K^{+}}(z)$,
and $D_{d}^{K^{+}}(z)$. There should be only three independent
curves without the difference between $\beta_{u}$ and $\beta_{s}$,
e.g., $D_{s}^{K^{+}}(z)$ equals to $D_{d}^{K^{+}}(z)$ as in the
original model. In principle, the flavor structure of the kaon
fragmentation functions is more complicated than that of the pion,
as revealed by the Field-Feynman model. However, some available
parametrizations \cite{BKKK,BKKK0,DK1,KKP} usually make only
distinction between the favored fragmentation functions, which are
related to the valence quarks of the kaon, and the unfavored
fragmentation functions, which are related to the light-flavor sea
quarks of the kaon.

\begin{figure}[htb]
\resizebox{0.5\textwidth}{!}{%
  \includegraphics{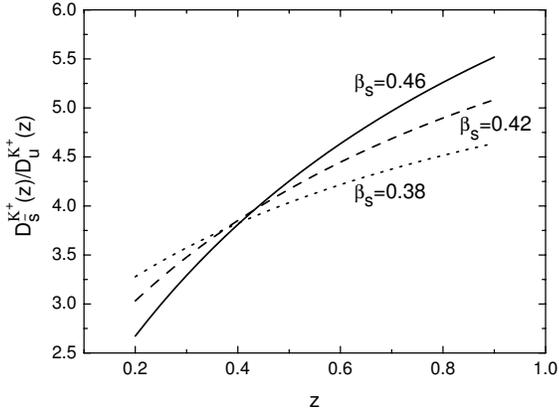}}
\caption{\small These three curves give the ratio between the
favored fragmentation functions $D_{\bar{s}}^{K^{+}}(z)/
D_{u}^{K^{+}}(z)$ in this work. It goes back to the original
Field-Feynman model when $\beta_{s}=\beta_{u}=0.46$. }\label{f9}
\end{figure}

\begin{figure}[ht]
\resizebox{0.5\textwidth}{!}{%
  \includegraphics{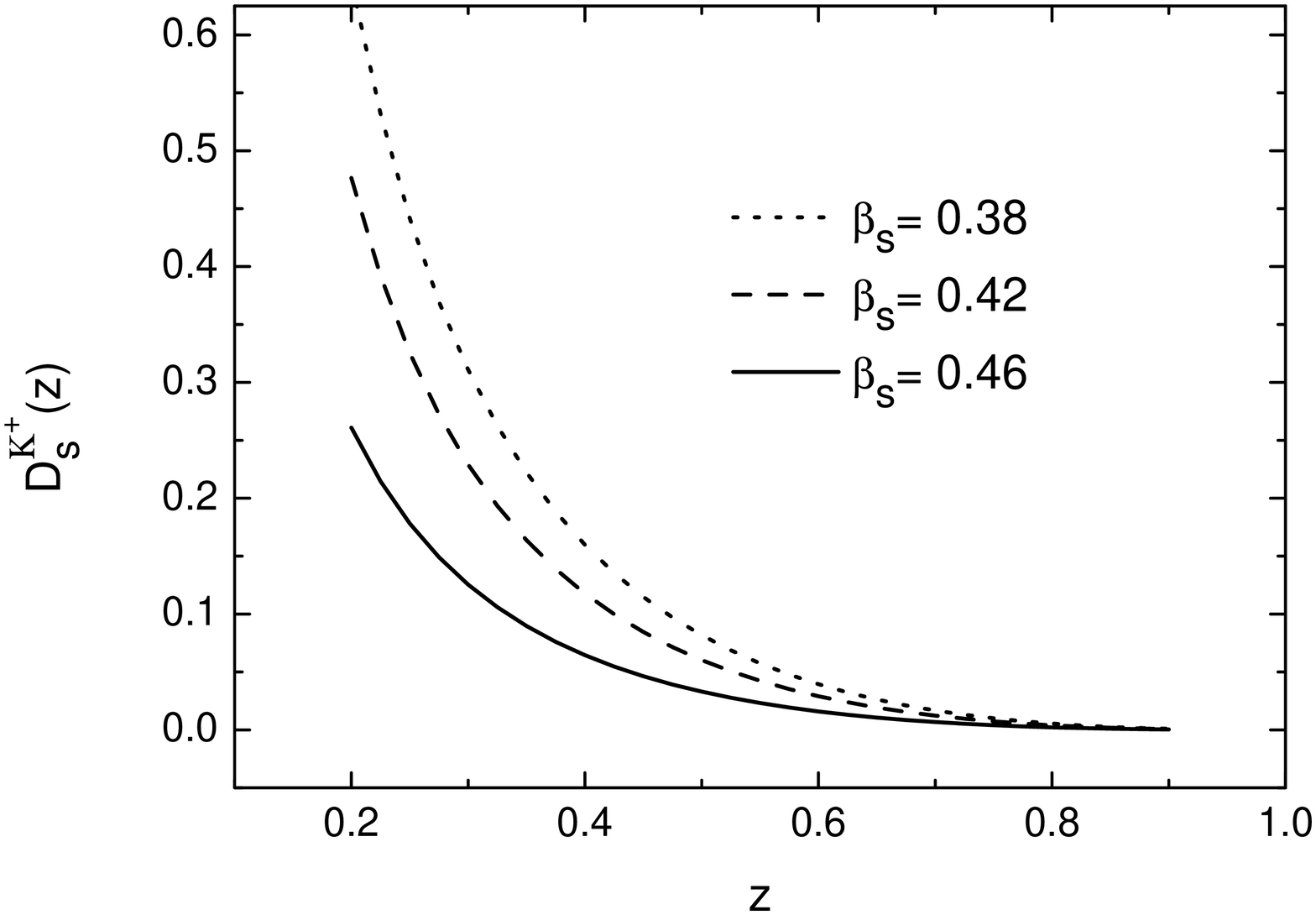}}
\vspace{0.05cm} \caption{\small These three curves show the
unfavored fragmentation function $D_{s}^{K^{+}}(z)$ with different
$\beta_{s}$ in this work. It goes back to the original
Field-Feynman model when $\beta_{s}=\beta_{u}=0.46$. }\label{f10}
\end{figure}

The effect of introducing $\beta_{s}\neq \beta_u$ might be small
in the fragmentation functions of the pion, but it is amplified in
the fragmentation functions of the kaon. It does not only
introduce the difference between the unfavored fragmentation
functions $D_{s}^{K^{+}}(z)$ and $D_{d}^{K^{+}}(z)$, but also
causes modification to the favored fragmentation functions
$D_{\bar{s}}^{K^{+}}(z)$.  In Fig.~\ref{f9}, we plot the ratio of
the favored fragmentation functions:
$D_{\bar{s}}^{K^{+}}(z)/D_{u}^{K^{+}}(z)$. The curve with
$\beta_{s}=0.42$ is adopted as an alternative option for
comparison, and the effect of $\beta_{s}$ on the unfavored
fragmentation function is illustrated in Fig.~\ref{f10}. The curve
with $\beta_{s}=0.46$ is the result in the original model, where
$\beta_{s}=\beta_{u}=0.46$. Therefore we notice a significant
modification to the favored fragmentation function
$D_{\bar{s}}^{K^{+}}(z)$ in comparison with that in the original
model. We also plot the prediction of the unfavored fragmentation
function $D_{s}^{K^{+}}(z)$, and find that it differs from
$D_{d}^{K^{+}}(z)$ significantly. These predictions can be tested
by future experiments, such as the semi-inclusive kaon productions
in deep inelastic scattering by the HERMES collaboration and the
CLAS collaboration {\it et al.}. The parameters used in this work
can be also constrained by further detailed studies, or the
framework should be extended to more general situation with more
sophisticated aspects taken into account.

\section{Summary}

In summary, we analyzed the fragmentation functions of both the
pion and kaon in the Field-Feynman recursive model with
significant modification. By taking into account the difference
between the excitation of sea quark-antiquark pairs by the initial
light-flavor ($u$ or $d$) and strange ($s$) quarks, we can
distinguish between $D_{u}^{\pi^{+}}$, $D_{d}^{\pi^{+}}$, and
$D_{s}^{\pi^{+}}$ in the Field-Feynman model with only three
parameters. The analytical results obtained are compatible with
available empirical results, and can be served as an alternative
set of completed fragmentation functions of the pion. We also
extended the same framework to the kaon, and provided a new set of
fragmentation functions of the kaon with clear distinction between
$D_{\bar{s}}^{K^{+}}(z)$, $D_{u}^{K^{+}}(z)$, $D_{s}^{K^{+}}(z)$,
and $D_{d}^{K^{+}}(z)$. Thus this work provides the fragmentation
functions of the pion and kaon with a same set of only three
parameters. The predictions can be tested by future experiments on
the fragmentation functions of the pion and kaon.

\vspace{0.5cm}

\noindent {\it Acknowledgments:} This work is partially supported
by National Natural Science Foundation of China under Grant
Numbers 10025523 and 90103007.

\vskip 1cm


\end{document}